\documentclass[twocolumn]{aastex62}
\usepackage{chngcntr}
\usepackage{booktabs}

\counterwithout{figure}{section}

\graphicspath{ {./figures/} }
\submitjournal{ApJ}

\shorttitle{Time-dependence of GI spirals}
\shortauthors{Hall et al.}

\begin{document}

\title{The temporal requirements of directly observing self-gravitating spiral waves in protoplanetary discs with ALMA}

\correspondingauthor{Cassandra Hall}
\email{cassandra.hall@le.ac.uk}

\author{Cassandra Hall}
\affil{Dept. of Physics \& Astronomy, University of Leicester, University Road, Leicester, LE1 7RH, U.K.}

\author{Ruobing Dong}
\affiliation{Department of Physics \& Astronomy, University of Victoria, Victoria BC V8P 1A1, Canada}

\author{Ken Rice}
\affiliation{SUPA, Institute for Astronomy, Royal Observatory Edinburgh, University of Edinburgh, Blackford Hill, Edinburgh, EH9 3HJ, UK}
\affiliation{Centre for Exoplanet Science, University of Edinburgh, Edinburgh, U.K.}

\author{Tim J. Harries}
\affiliation{Department of Physics and Astronomy, University of Exeter, Stocker Road, Exeter EX4 4QL, U. K.}

\author{Joan Najita}
\affiliation{National Optical Astronomy Observatory, 950 N. Cherry Avenue, Tucson, AZ 85719, USA}

\author{Richard Alexander}
\affiliation{Dept. of Physics \& Astronomy, University of Leicester, University Road, Leicester, LE1 7RH, U.K.}

\author{Sean Brittain}
\affiliation{Department of Physics \& Astronomy, 118 Kinard Laboratory, Clemson University, Clemson, SC 29634-0978, USA}


\begin{abstract}
We investigate how the detectability of signatures of self-gravity in a protoplanetary disc depends on its temporal evolution. We run a one-dimensional model for secular timescales to follow the disc mass as a function of time. We then combine this with three-dimensional global hydrodynamics simulations that employ a hybrid radiative transfer method to approximate realistic heating and cooling. We simulate ALMA continuum observations of these systems, and find that structures induced by the gravitational instability (GI) are readily detectable when $q=M_\mathrm{disc}/M_*\gtrsim 0.25$ and $R_\mathrm{outer}\lesssim 100$ au. The high accretion rate generated by gravito-turbulence in such a massive disc drains its mass to below the detection threshold in $\sim10^4$ years, or approximately 1\% of the typical disc lifetime. Therefore, discs with spiral arms detected in ALMA dust observations, if generated by self-gravity, must either be still receiving infall to maintain a high $q$ value, or have just emerged from their natal envelope. 
Detection of substructure in systems with lower $q$ is possible, but would require a specialist integration with the most extended configuration over several days. This disfavours the possibility of GI-caused spiral structure in systems with $q<0.25$ being detected in relatively short integration times, such as those found in the DSHARP ALMA survey \citep{dsharp1,dsharpspirals}. We find no temporal dependence of detectability on dynamical timescales.\\~\\
\end{abstract}

\section{Introduction} \label{sec:intro}

In the last decade, huge leaps in imaging capabilities have allowed astronomers to obtain high-resolution images of protoplanetary discs, the birth sites of exoplanets. Among them, near infrared (NIR) imaging allows us to probe the surface of discs (Fukagawa et al. 2013),
while dust continuum observations at $\sim$mm wavelengths carried out using the Atacama Large Millimeter/submillimeter array (ALMA) probe down to the midplane of discs and trace the density structures in $\sim$mm-sized dust. Surprisingly, a significant fraction of these discs have substructures, such as rings \citep[e.g.,][]{hltau} and spirals \citep[e.g.,][]{perezetal2016}, and these substructures seem to be common; images of discs from surveys that are spatially resolved down to 28 au show substructure in about 20\% of these objects \citep{ALMAoph2018}.



It is widely thought that rings may be caused by planetary-mass companions \citep{kleynelson2012,baruteauetal2014,dipierrohltau,dongzhuwhitney2015,dongfung2017,dipierroetal2018}. However, at present, we lack the data to distinguish between their formation through planet-disc interactions and other possible mechanisms, such as self-induced dust pileups \citep{gonzalezlaibe2015} or aggregate sintering \citep{okuzumietal2016}. The origin of spiral features in protoplanetary discs is just as murky \citep{dongetal2018}. Although planets can induce spiral features in scattered light images \citep{dongzhu2015}, it is unclear if they can do so in $\sim$mm emission that trace the distribution of $\sim$mm-sized dust. 

A possible explanation for spirals present in $\sim$mm emission is the gravitational instability (GI). At the moment of formation of a star-disc system, the masses of the star and the disc are comparable, guaranteeing that the system is self-gravitating \citep{linpringle1987,linpringle1990}. This ensures that the Toomre parameter, $Q$, of such a system \citep{toomre1964}
\begin{equation}
\label{eq:toomre}
Q=\frac{c_\mathrm{s}\kappa}{\pi\mathrm{G}\Sigma} \sim 1,
\end{equation}
where $c_\mathrm{s}$ is the sound speed, $\kappa$ is the epicyclic frequency ($\kappa=\Omega=\sqrt{\mathrm{G}M/r^3}$ in a Keplerian disc) and $\Sigma$ is the surface density. So long as $Q~\sim 1.5-1.7$, numerical simulations have shown that non-axisymmetric perturbations will grow into spiral waves \citep{durisenetal2007}. If the disc is able to cool rapidly relative to the dynamical timescale, the spiral arms may then fragment \citep{gammie2001,rice2005,stamatellos2007,kratter2010failedbinary,nayakshin2010,forganricepopsynth2013,halletal2017,forganpopsynth2018,stamatellosinutsuka2018,humphriesnayakshin2018}. 

Equation \ref{eq:toomre} is, however, a local condition for instability. Since observations usually give us global properties of a system, it is useful to think about the global requirement for instability, which is simply that the disc-to-star mass ratio,
\begin{equation}
q\equiv\frac{M_\mathrm{disc}}{M_*}=f\cdot \frac{H}{R} \gtrsim 0.1,
\end{equation}
\citep{kratterlodato2016}, where $M_\mathrm{d}$ and $M_*$ are the mass of the disc and the star respectively, $H=c_\mathrm{s}/\Omega$ is the disc scale height, and $f$ is a numerical prefactor of order unity.

Understanding gravitational instability, and subsequent fragmentation, requires observations of protoplanetary discs that are likely to be gravitationally unstable. It has been suggested that some systems with spiral arms are gravitationally unstable. For example, the grand design, $m=2$ spiral modes imaged in scattered light in MWC 758 \citep{benistyetal2015} and SAO 206462 \citep{stolkeretal2016} are consistent with spirals in gravitational instability models \citep{donghallrice2015}. On the other hand, the difference between the $q\gtrsim0.1$ required for the disc to be self-gravitating, and the $q\sim0.01$ estimated from $\sim$mm dust emission \citep{andrews2011} leaves this scenario unfavored. However, if part of the disc is optically thick at $\sim$mm wavelengths, disc mass could be significantly underestimated \citep{hartmann2006,forganrice2013,dunhametal2014,evansetal2017,galvanmadridetal2018}. 

ALMA has recently revealed spiral arms in $\sim$mm emission in many systems \citep{tobin2016,perezetal2016,dongliuetal2018}. The first of its kind, the Elias 2-27 system, is a class II object with an unusually high mm emission-based disc mass estimate -- $q\sim 0.24$ \citep{andrewsetal2009}. It has a two-armed, grand-design spiral extending out to $R\sim 300$ au from the central star \citep{perezetal2016}. Both GI and an external perturber have been put forward to explain the origin of the spirals \citep{meruetal2017, tomidaetal2017}, and efforts to distinguish the two are ongoing  \citep{forganetal2018}.

Some previous investigations into spirals detected in ALMA continuum observations have suggested that we should be cautious about assuming they are due to GI.
Even if a disc has $q > 0.1$, its GI-induced structures are not necessarily detectable, since their amplitudes may not be large enough \citep{halletal2016}. Such features may also be smeared so that their apparent morphology is different to their actual morphology.  For example, spiral arms may be smeared into $\sim 2$ when $\sim8-10$ are actually present \citep{dipierro2014,dipierro2015}. Similarly, although GI models could explain the morphology of the Elias 2-27 system \citep{meruetal2017}, fine tuning of the parameter space is needed \citep{halletal2018}, as the extended nature of the disc may make it susceptible to fragmentation (e.g., \citealt{rafikov2005}). 

Spiral arms, particularly grand-design two-armed ones, are being revealed as common (up to $\sim$20\%) in high resolution imaging surveys in both NIR scattered light \citep{dongetal2018} and mm continuum observations \citep{ALMAoph2018, dsharp1,dsharpspirals,kurtovicdsharp}. It is therefore of critical importance to determine the physical mechanism, or possible mechanisms, that are driving them. Unfortunately, the two most widely considered scenarios, companion and GI, are both difficult to verify in individual systems. To confirm the former, direct imaging observations searching for companions are needed. Such observations are challenging (e.g., \citealt{testietal2015,maireetal2017}), particularly if planets form in the ``cold start'' instead of the commonly assumed  ``hot start'' scenario
\citep{spiegelburrows2012}. Therefore, except in rare cases (e.g., HD 100453, \citealt{dongetal2016,wagneretal2018}), arm-driving companions have not been confirmed. 

To verify GI as the arm-driving mechanism, accurate measurements of the total mass of the gas disc are required. The most common avenue of estimating disc masses from sub-mm dust continuum relies on knowledge of the dust-to-gas mass ratio and the optical properties of dust grains \citep{beckwith1990}, both of which are highly uncertain, and can lead to underestimating the disc masses by a factor of up to $\sim 100$ \citep{forganrice2013}. Estimating disc masses through $^{13}$CO and C$^{18}$O isotopolgues emission is possible \citep{williamsbest2014}, but this method is model dependent and suffers from uncertainties in, for example, the chemistry of CO \citep{ileeetal2011,ileeetal2017,  yuetal2017}.

Given the difficulties in determining the origin of observed spiral arms directly and in individual systems, we explore the likelihoood of observing GI-induced spiral arms in discs as a sample. We follow the time evolution of an isolated disc that has just emerged from its natal envelope to an age of $\sim 10$ Myr. The system undergoes angular momentum transport primarily due to the gravitational instability. Our goal is to take a holistic approach. Rather than attempting to explain the morphologies of individual systems, we ask a broader question. Given its observability, and the observed occurrence rate of spirals in discs, how likely is it that GI is the dominant spiral-driving mechanism in protoplanetary discs?

\section{Method}

We begin with a one-dimensional model of an evolving self-gravitating disc, and use this to obtain disc masses at times that are representative of evolutionary stages in the paradigm of evolving protoplanetary discs. We use this model to set the disc-to-star mass ratios of global, three-dimensional hydrodynamics simulations of self-gravitating discs at representative epochs. Once evolved for a few orbital periods at the disc outer edge, we perform radiative transfer calculations and generate synthetic images to predict how such systems would be observed by ALMA.  

The disc is modelled in isolation. We define time as $t_\mathrm{tot}=t_0 + t $, where $t_\mathrm{tot}$ is the total time (i.e., system age), $t_0$ is the point at which disc accretion dominates over infall, and $t$ refers to simulation time (throughout the paper, ``time'' refers to $t$). Most likely, $t_0$ occurs during the late stages of Class 0 or early Class I phase, while a partial envelope may still be present.
Prior to $t_0$, we can crudely think of the envelope as supplying mass at a constant rate to the disc, such that the disc maintains a time independent surface density profile, and therefore constant total mass. After $t_0$, envelope infall has effectively ceased, disc accretion continues, draining the disc onto the central protostar.

\subsection{Time-dependent one dimensional model}
\begin{figure}
  \includegraphics[width=\linewidth]{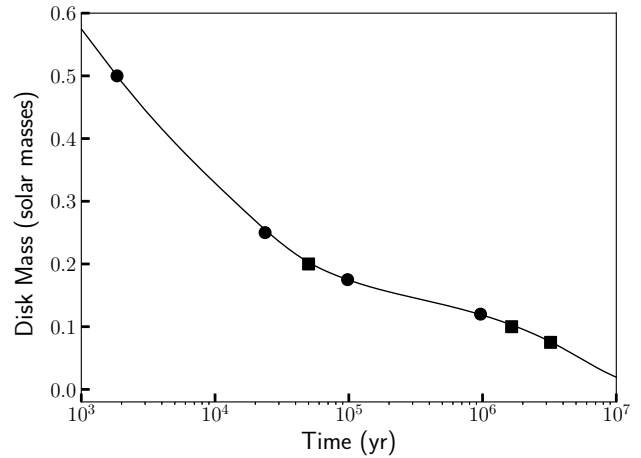}
  \caption{The time-dependent mass evolution of a self-gravitating disc orbiting a 1 M$_\odot$ star, obtained self-consistently using the models of \citet{ricearmitage2009}. The line shows the disc mass as a function of time, with markers showing the points that were used to set the disc masses of the SPH simulations. Circular markers show points in time used for SPH simulations and synthetic observations. Square markers show points in time used only for SPH simulations. 
 \label{fig:1D}}
  \end{figure}

We use the one-dimensional model of \citet{ricearmitage2009}, which evolves a self-gravitating protoplanetary disc under the assumption that the gravitational potential is fixed, angular momentum transport is primarily due to disc self-gravity, and the disc is in thermal equilibrium. The model does, however, assume that there is a minimum viscous $\alpha$, which could be produced via, e.g., the magnetorotational instability. It also includes a disc wind that dominates, and dissipates the disc mass, when the mass accretion rate is low.  Full details are given in \cite{ricearmitage2009}.  However, we outline the basics of the model here.

Since we assume that the disc evolves pseudo-viscously, the surface density, $\Sigma(r,t)$, evolves according to \citep{bellpringle1974, pringle1981}
\begin{equation}
\label{eq:dsigma/dt}
\frac{\partial \Sigma}{\partial t} = \frac{3}{r}\frac{\partial}{\partial r} \Bigg[r^{\frac{1}{2}} \frac{\partial}{\partial r}  \bigg(\nu\Sigma r^{\frac{1}{2}}\bigg)  \Bigg]
\end{equation}
where $\nu$ is the kinematic viscosity. We can express this viscosity as $\nu = \alpha c_s H$, where $\alpha \ll 1$ is the viscosity parameter \citep{shakurasunyaev1973}.
If the disc can maintain a quasi-steady state, equation \ref{eq:dsigma/dt} can be integrated to give the steady-state mass accretion rate \citep{pringle1981}
\begin{equation}
\dot{M} = 3\pi\nu\Sigma.
\end{equation}
Viscosity generates dissipation in the disc at the rate \citep{belllin1994}
\begin{equation}
D(R) = \frac{9}{4}\nu\Sigma\Omega^2,
\end{equation} 
where $D(R)$ is per unit area per unit time. 
Assuming a quasi-steady state, heating is balanced by cooling, with the cooling time, $t_\mathrm{cool}$, given by  \citep{gammie2001}
\begin{equation}
t_\mathrm{cool}=\frac{4}{9\gamma(\gamma -1)\alpha\Omega},
\end{equation}
where $\gamma$ is the ratio of specific heats. The cooling time, $t_\mathrm{cool}$, can also be expressed as $t_\mathrm{cool}=U/\Lambda$, where $U$ is the internal energy per unit area,
\begin{equation}
U=\frac{c_s^2 \Sigma}{\gamma(\gamma -1)},
\end{equation}
and the cooling rate $\Lambda$ is given by \citep{pringle1981,hubeny1990,johnsongammie2003,ricearmitage2009}
\begin{equation}
\Lambda = \frac{16\sigma}{3}(T_\mathrm{mid}^4 - T_0^4) \frac{\tau}{1+\tau^2}.
\end{equation}
Here, $T_0=3$ K and is assumed to come from a background irradiation source that prevents the midplane of the disc from cooling below this value  \citep{stamatellos2007}.  The optical depth is approximated using 
\begin{equation}
\label{eq:tau}
\tau = \int^{\infty}_{0} \mathrm{d}z\kappa(\rho_z,T_z)\rho_z\approx H\kappa(\bar{\rho},\bar{T})\bar{\rho},
\end{equation}
where $\kappa$ is the Rosseland mean opacity (obtained from \citealt{belllin1994}), $\bar{\rho}=\Sigma/(2H)$, and $\bar{T}=T_\mathrm{mid}$. 

Closing equations \ref{eq:dsigma/dt} to \ref{eq:tau} is our assumption that the disc will settle into a self-gravitating state with $Q=1.5$. Given the surface density, we are therefore able to estimate the sound speed, the cooling timescale, the equilibrium heating rate and, hence, $\alpha$. In our model, we do not include that some of the mass flowing through the disc will accrete onto the central star, and increase its mass. We instead assume that mass accreted through the disc is completely lost in a jet. This does mean that our model is a simplification, however, most of the mass is accreted within the first $10^5$ years \citep{ricemayoarmitage2010}, so this will not significantly change the relationship between system age and strength of the gravitational instability. We are, essentially, considering the best case scenario, where $q$ remains as high as possible for as long as possible.

\subsection{SPH simulations and emission maps}


The smoothed particle hydrodynamics (SPH) simulations \citep{gingoldmonaghan1977,lucy1977} are based on the code developed by \citet{bateetal1995}, updated to include a hybrid radiative transfer method that approximates realistic heating and cooling \citep{forgan2009,stamatelloswhitworth2009}. 
Essentially, the polytropic cooling approximation of \citet{stamatellos2007} is combined with the flux-limited diffusion method of \citet{mayeretal2007}, which together can account for the local optical depth of the system as well as the energy exchange between particles. 

The disc is heated through $P\,\mathrm{d}V$ work, and we assume the central star mass is 1 M$_\odot$. We run 7 models, with disc-to-star mass ratios of $q=0.5,0.25,0.2,0.175, 0.125,$ $0.1$ and $0.075$.

Each disc has $5\times 10^5 $ particles, initially located between 6 and 60 au, and the initial surface density profile and initial sound speed profile are $\Sigma\propto r^{-1}$ and $c_s \propto r^{-1/4}$.


We use the \texttt{TORUS} radiation transport code \citep{harriesetal2004,kurosawaetal2004,haworthetal2015} to calculate continuum emission maps of the SPH discs using the dust temperatures directly from the SPH simulations.
To do so, a 3D grid must be constructed from the particle distribution. Full details of this are given in \cite{rundleetal2010}, but the basic idea is to begin with one cell centered on the entire disc, and then to repeatedly divide this cell according to a resolution criterion (for example, resolve $n$ particles per cell). The original cell is divided once in each dimension, resulting in $2^D$ child cells, where $D$ is number of dimensions. We resolve the mass represented by every active particle on the grid, resulting in $\sim 500,000$ grid cells.

The dust in our model is \cite{drainelee} silicates, with a grain size distribution given by 
\begin{equation}
n(a) \propto a^{-q}  \hspace{1cm}\mathrm{for}\hspace{1cm} a_{\mathrm{min}} < a < a_{\mathrm{max}},
\end{equation}
where $a_\mathrm{min}$ and $a_\mathrm{max}$ are the minimum and maximum grain sizes, taken to be 0.1 $\mu$m and $2000$ $\mu$m respectively, and $q=3.5$, the standard power-law exponent for the ISM \citep{grainsize}. We assume a dust-to-gas ratio of 1:100 everywhere in the disc \citep{meruetal2017,tomidaetal2017}.
Previous numerical work has shown that it is possible to increase the fraction of grains present in the spiral arm of a self-gravitating disc through particle trapping \citep{riceetal2004}. Regardless, we do not expect this effect to significantly affect $\sim$mm grains. We do, however, discuss the implications of our assumptions in the Summary and Discussion section.

%

\subsection{Detecting substructure in synthetic ALMA images}
\label{sec:ALMA}

\begin{table}[]
\centering
\begin{tabular}{@{}ccccc@{}}
\toprule
 t (yr)& q & int.        & beam  & PWV \\ 
        &    & time (hrs) & size  &   (mm)        \\ \midrule 
$\sim$$10^3$ & 0.5     & 1     & $0.1\arcsec\times 0.09\arcsec$   & 1.796\\
$\sim$$10^4$ & 0.25   & 12   & $0.12\arcsec\times 0.08\arcsec$ & 1.796\\
$\sim$$10^5$ & 0.175 & 72   & $0.05\arcsec\times 0.04\arcsec$  & 0.45\\
$\sim$$10^6$ & 0.125 &  120  &$0.03\arcsec \times 0.02\arcsec$ & 0.45 \\ 
\bottomrule
\end{tabular}
\caption{Parameters used to generate synthetic observations. From left to right is simulation time in years, disc-to-star mass ratio, total integration time on source in hours, beam size in arc seconds and precipitable water vapour in mm.}
\label{tab:obs}
\end{table}

The emission maps generated by TORUS are used as inputs to the ALMA simulator included in \texttt{CASA} (ver 4.7.2) \citep{CASA}, and all discs were imaged at a distance of 139 pc, as if in the $\rho$-Ophiuchus region \citep{mamajek2008}. We synthesise observations centered on 230 GHz (band 6), chosen such that there is a balance between the disc being more optically thin (where longer wavelengths are preferred), and obtaining a higher signal-to-noise ratio (where shorter wavelengths are preferred). We choose the maximum bandwidth available in ALMA band 6 (7.5 GHz), since this maximises sensitivity.  We corrupt the visibilities with thermal noise by using the Atmospheric Transmission at Microwaves (\texttt{ATM}) code \citep{ATMcode}. The total integration time, beam sizes and precipitable water vapour (PWV) values are given in Table \ref{tab:obs}.

For the $q=0.5$ and $q=0.25$ discs, PWV values are chosen as an estimate from the ALMA sensitivity calculator at the fifth octile for this wavelength. We assume exceptional observing conditions in the case of $q=0.175$ and $q=0.125$, motivated by PWV $<$ 0.7mm 50\% of the time in August. We use the \texttt{simobserve} and \texttt{simanalyze} routines in \texttt{CASA}, which perform a “standard” clean, using Briggs weighting of the visibilities with a robust parameter of 0 and multi scale deconvolution \citep{raucornwell2011}.

We generate synthetic images at 4 stages of the disc lifetime, when $q=0.5$ ($t\sim 10^3$ yr), $q=0.25$ ($t\sim 10^4$ yr), $q=0.175$ ($t\sim 10^5$ yr) and $q=0.125$ ($t\sim 10^6$ yr). We begin with a shorter integration time on the most massive disc (1 hour), since the most massive disc has the largest total flux and so smaller integration times will suffice. As the disc mass decreases, we use progressively longer integration times. Essentially, the observing parameters were varied in order to maximise the detection of spiral arms in the shortest integration time, so the images presented here show the most clear results with optimal use of resources.

In all cases, the surface density profile of a self-gravitating disc in quasi-steady equilibrium is steep, which makes it difficult to observe the fainter non-axisymmetric structure away from the center of the disc. To reduce the overall range in the image (which enhances the fainter features), we convolve each image with a 2D Gaussian, and then subtract this from the original image to obtain the ``residuals'' (i.e., the unsharped image masking operation; \citealt{unsharp}; see applications in, e.g., \citealt{perezetal2016}). \\~\\

\section{Results}

\subsection{Temporal behaviour on secular timescales}
\label{subsec:secular}
\begin{figure*}
\begin{tabular}{cccc}
\hspace{1.2cm} $t\sim10^3$yr & \hspace{1.2cm} $t\sim 10^4$ yr & \hspace{1.2cm} $t\sim 10^5$ yr & \hspace{1.3cm}$t\sim 10^6$ yr \\
\hspace{1.2cm}($M_\mathrm{d}=0.5$ M$_\odot$) & \hspace{1.5cm}($M_\mathrm{d}=0.25$ M$_\odot$) &\hspace{1.6cm}  ($M_\mathrm{d}=0.175$ M$_\odot$) &  \hspace{1.4cm}($M_\mathrm{d}=0.125$ M$_\odot$)  \\
\end{tabular}\\
\rotatebox{90}{\hspace{0.3cm}-100 \hspace{1cm} 0 \hspace{1cm} 100}\hspace{2mm}\includegraphics[width=\textwidth]{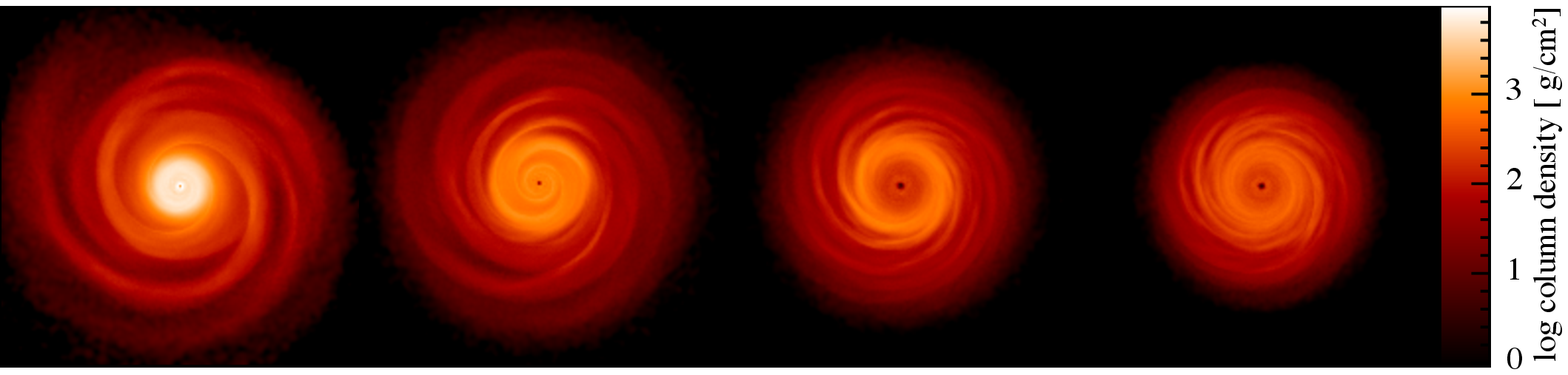}
     \hspace*{-1.5cm}\includegraphics[clip,trim={ 2.5cm 0.5cm 0 1.2cm},width=1.2\textwidth,height=0.225\textheight]{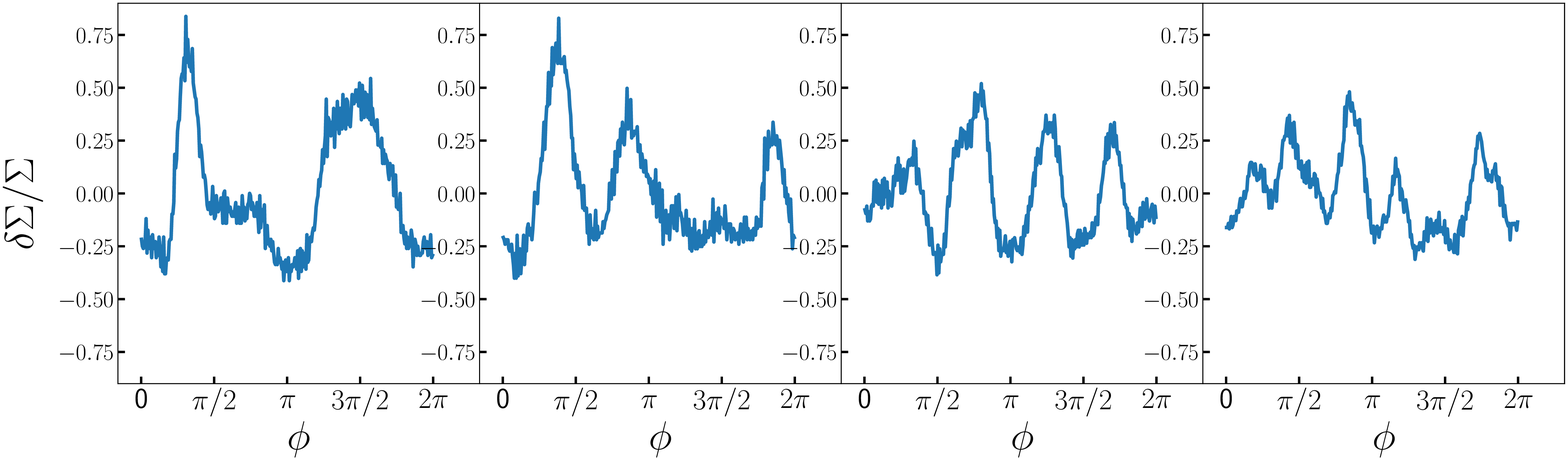}\\
\begin{tabular}{cccc}
\hspace{2cm}1 hour & \hspace{3cm}12 hours & \hspace{2.5cm} 72 hours & \hspace{3cm}120 hours\\
\end{tabular}\\
  \hspace*{-1cm}\includegraphics[clip,trim={1cm 2cm 1cm 2.5cm},width=1.1\textwidth]{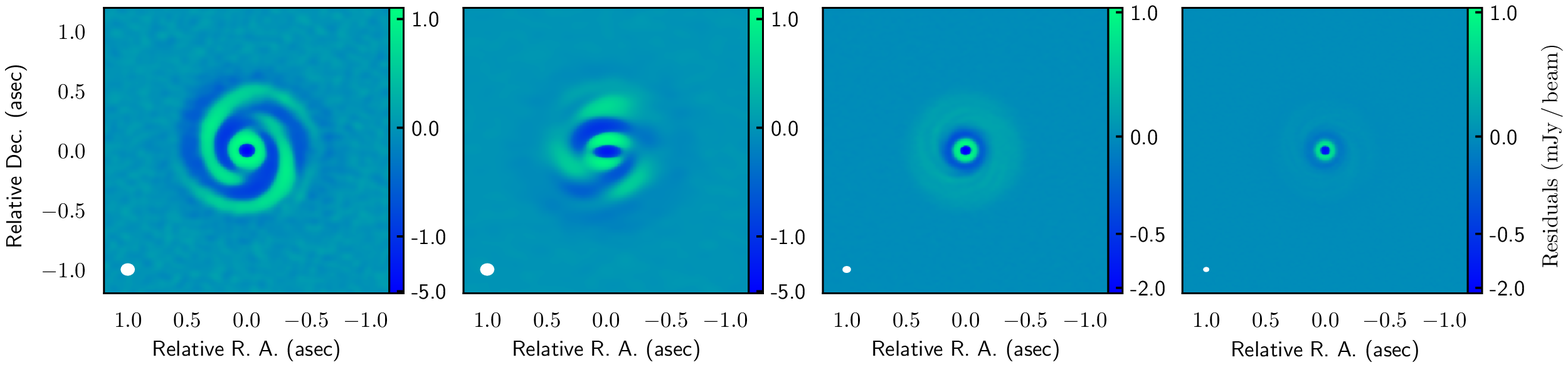}
  \hspace*{-1cm}\includegraphics[clip,trim={ 2cm 0 0 1.2cm},width=1.17\textwidth,height=0.225\textheight]{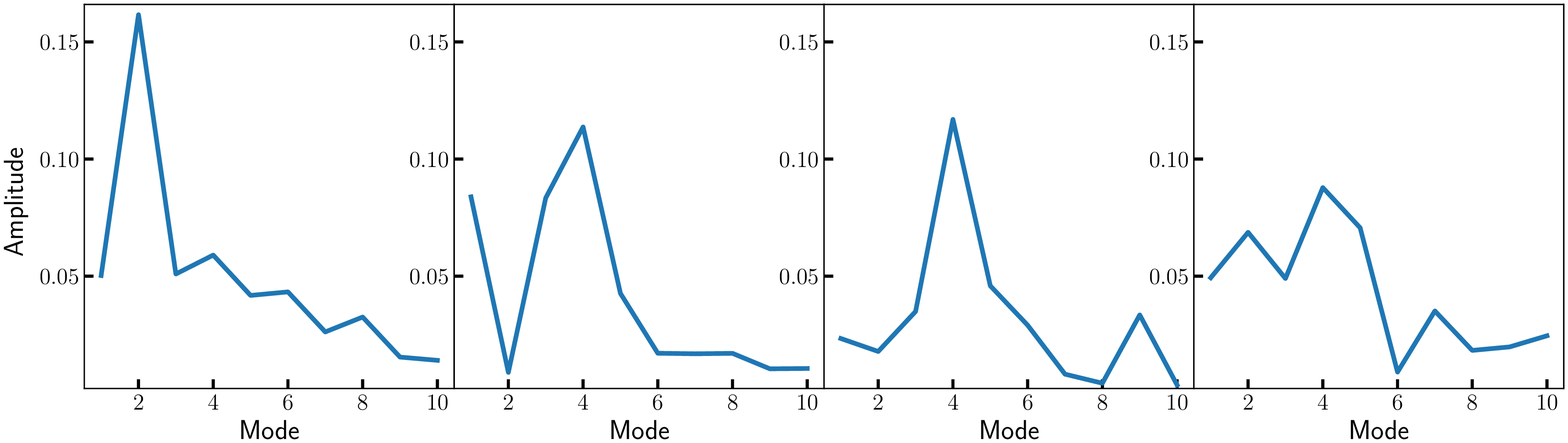}

  \caption{Top row: Surface density structure of the discs considered. Second row: Azimuthal surface density variation between $R=40$ au and $R=45$ au (equation 11). Third row: Residuals of synthetic ALMA observations that have been convolved with a 2D Gaussian and then subtracted from the original (section \ref{sec:ALMA}). This is to enhance the fainter substructure in the outer parts of the disc. Non-axisymmetric structure becomes increasingly difficult to detect as 1) spiral amplitudes decrease 2) dominant $m$-mode increases. Bottom row: Fourier amplitude of each disc computed in a ring between $R=40$ au and $R=45$ au. Non-axisymmetric structure is only visible in the ALMA residuals when there is sufficient power in the low $m$-modes.
  \label{fig:multipanel}}
\end{figure*}  

\begin{figure}
\includegraphics[width=1.4\linewidth,trim={0 70 0 0}]{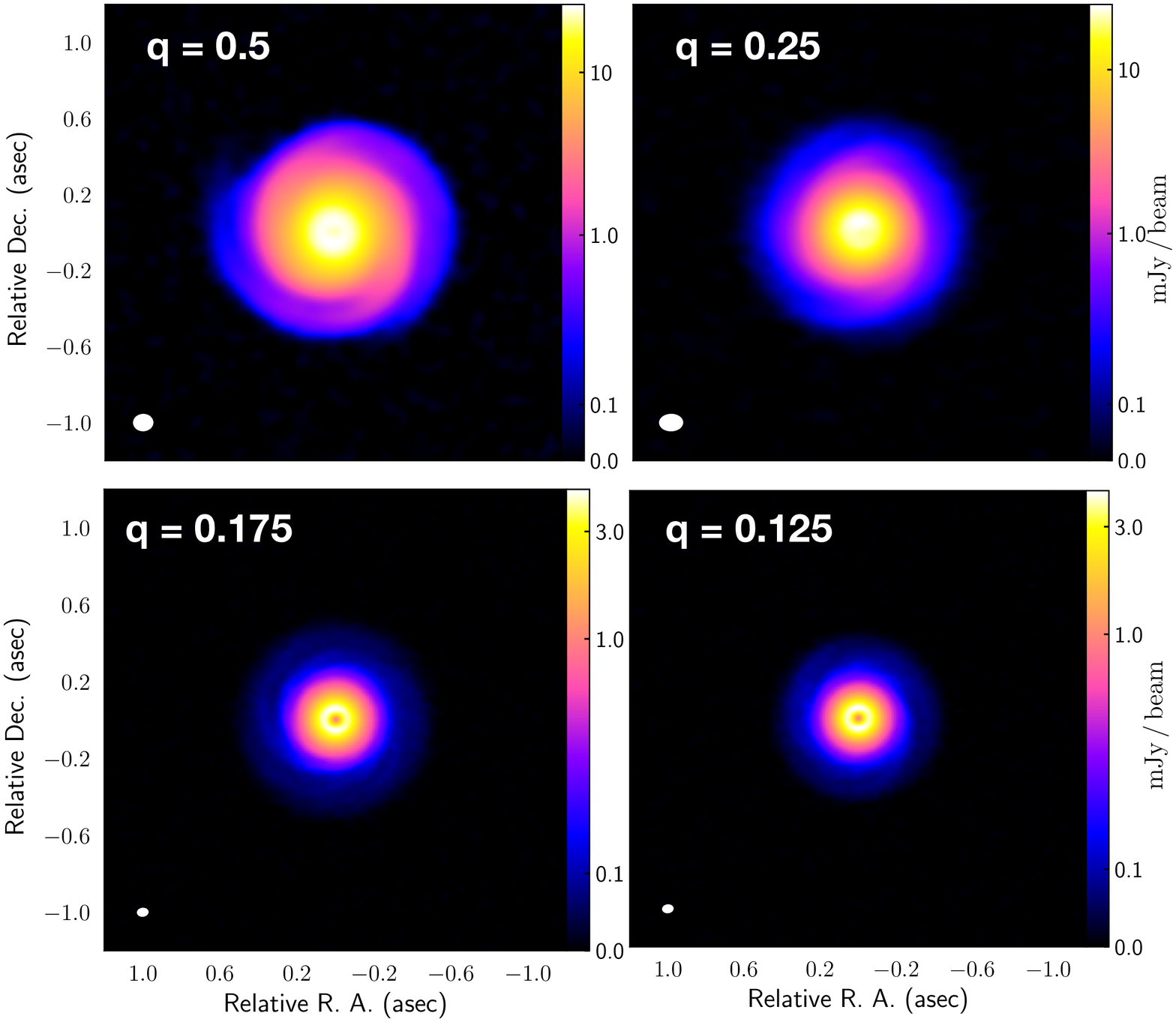} 
\caption{Top left: synthetic ALMA images of the $M_\mathrm{d}=0.5$ M$_\odot$ disc, with a peak flux of 26.15 mJy beam$^{-1}$.  Top right: synthetic ALMA image of the $M_\mathrm{d}=0.25$ M$_\odot$ disc, with a peak flux of 25.04 mJy beam$^{-1}$. Bottom left: synthetic ALMA images of the $M_\mathrm{d}=0.175$ M$_\odot$ disc, with a peak flux of 4.62 mJy beam$^{-1}$. Bottom right: synthetic ALMA images of the $M_\mathrm{d}=0.125$ M$_\odot$, with a peak flux of 4.33 mJy beam$^{-1}$.  \label{fig:uncorrected}}
\end{figure}

We ran 7 global SPH simulations, with initial disc masses taken from 7 points in time from Figure \ref{fig:1D}. Doing so allowed us to capture the secular behaviour of the system, while simulating for several outer orbits allowed us to capture behaviour occurring on dynamical timescales. While the 1D models of \citet{ricearmitage2009} provide surface density profiles of the quasi-steady state discs (expected roughly after the thermal timescale of the system), we did not use them for the initial conditions. This is because the azimuthally averaged surface density profiles of the 3D simulations are slightly different due to capturing the non-axisymmetric structure of the system (i.e., large-scale, global spiral arms). In both the 1D and the 3D case, the qualitative behaviour is the same. Beginning with some imposed surface density profile, the system evolves to a surface density profile in the quasi-steady state that is independent of the initial configuration of the system.

We begin with the discs in a $\Sigma \propto r^{-1}$ configuration,
and allow them to evolve to their quasi-steady profile. The temperature profile is initially $T\propto r^{-1/2}$, with the temperature normalised such that the minimum value of $Q$ is $Q_\mathrm{min}=2$. This is a local parameter, so $Q=Q_\mathrm{min}$ only at the disc outer edge. The discs subsequently cool until $Q$ is low enough for gravitational instability to set in, which then provides heating. The heating and cooling ultimately roughly balance and the disc settles into a quasi-steady state. 


The top row of Figure \ref{fig:multipanel} shows the surface density structure for four simulations with different total disc masses. The second row of Figure \ref{fig:multipanel} shows the fractional physical amplitude of the spiral between $R=40$ au and $R=45$ au,
\begin{equation}
\frac{\delta\Sigma}{\Sigma}=\frac{\delta\Sigma(\phi)_{r=40-45\,\mathrm{au}}}{\Sigma_{r=40-45\,\mathrm{au}}} = \frac{\Sigma(\phi)_{r=40-45\,\mathrm{au}}-\Sigma_{r=40-45\,\mathrm{au}}}{\Sigma_{r=40-45\,\mathrm{au}}}
\end{equation}
where $\Sigma_{r=40-45\,\mathrm{au}}$ is the azimuthally averaged surface density in a ring between $R=40$ au and $R=45$ au. It is clearly shown that as the disc ``evolves'' in time, both the number of peaks increases and the physical amplitude of the spiral decreases.  

The third row shows the synthetic ALMA residuals (section \ref{sec:ALMA}), which display the non-axisymmetric structure in the disc that would be observed in a real ALMA observation. Figure \ref{fig:uncorrected} shows the synthetic images without the unsharp mask technique applied, showing the necessity of applying the unsharp mask in order to enhance the fainter features and reduce the overall range of the image.  The residuals clearly show that as the number of spiral arms increases, and their physical amplitude decreases, it becomes increasingly difficult to detect non-axisymmetric structure, due to a decrease in overall flux, a decrease in contrast between the arm and inter-arm region, and the smaller physical scale of the spirals. 

When the disc mass is large ($q= 0.5$), and the spirals are loosely wound, an hour on source is sufficient to detect the substructure, suggesting that future protoplanetary disc surveys that perform hour-long integrations should be able to detect substructure due to GI.

As the disc mass decreases, the spirals become increasingly faint and more tightly wound, requiring both a smaller beam size and a far longer integration time. When $q=0.25$, 12 hours on source is required. When disc mass is decreased further ($q= 0.175$), 72 hours is required on source to detect substructure, with a more compact beam ($0.05\arcsec\times 0.04\arcsec$), as well as a smaller amount of PWV in the atmosphere.  Observing such structure would require a dedicated integration on a deliberately targeted source, which makes the detection of GI in protoplanetary discs with disc-to-star mass ratios of less than 0.25 unlikely. Finally, for the least massive disc ($q=0.125$), no detection of substructure is visible even with an integration time of 120 hours.

This can also be understood in terms of the Fourier amplitude, which we calculate for each mode in a ring between $R=40$ au and $R=45$ au. This is representative of the majority of the disc, and a relatively thin ring is required to avoid the structure being averaged out. The results do not change if the location of the ring is varied.
The Fourier amplitude, $A_m$, of each mode, $m$, is given by
\begin{equation}
A_m = \Bigg| \sum^{N_{\mathrm{ring}}}_{i=1}  \frac{\mathrm{e}^{-im\phi_i}}{N_\mathrm{ring}}\Bigg|,
\end{equation}
where $N_\mathrm{ring}$ is the number of particles in each ring and $\phi_i$ is the azimuthal angle of the $i^{\mathrm{th}}$ particle. The Fourier amplitude of the first 10 modes are displayed in the bottom row of Figure \ref{fig:multipanel}. 

We can see that when the disc just emerges from its natal envelope and is at its most massive state (leftmost column), two grand-design global spirals are clearly visible, with most of the power in the $m=2$ Fourier mode. This is shown in the bottom left panel of Figure \ref{fig:multipanel}.

As the disc evolves, the amplitude of this mode decreases, and more power may be found in lower $m$-modes. For the $q=0.25$ disc, this results in some non-axisymmetric structure remaining visible in the residual image, but not as clearly as when the $m=2$ mode dominates. As the disc continues to evolve, less and less power is in the lower $m$-modes, resulting in increased difficulty of detection.

\begin{figure}
\includegraphics[width=\linewidth]{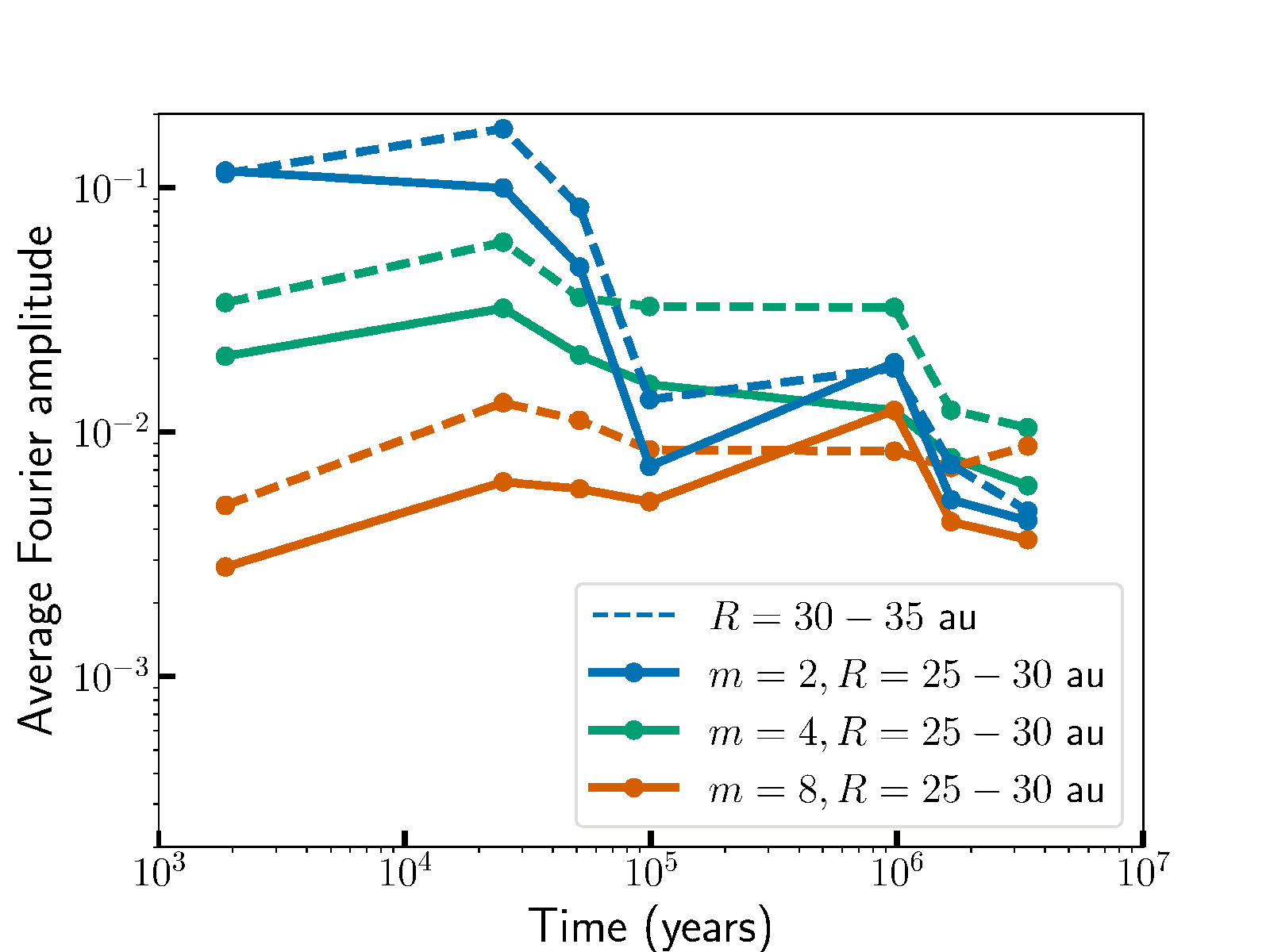}
\caption{Fourier amplitudes of even modes as a function of time. As time increases, the amount of power in the lower $m$-modes decreases, which makes detecting this spiral structure with an instrument such as ALMA increasingly difficult. As long as the amplitude remains above $\sim 0.1$ for a low $m$-mode (where $m\le 4$), GI spirals will be detectable by ALMA. \label{fig:avg}}
\end{figure}

In order to numerically quantify these results, we perform this Fourier analysis on 7 simulations, each representing the disc at a time taken from Figure \ref{fig:1D}. We take the Fourier amplitude of each mode, and average it over $\sim 4$ orbital periods at the disc outer edge. The results are shown in Figure \ref{fig:avg}. We can see that up until a few $\times10^4$ years ($\sim 100$ orbits at $R=100$ au), 
the $m=2$ mode dominates the spectrum.
From the two leftmost columns of Figure \ref{fig:multipanel}, we can see that GI-induced structure is detectable when 
(1) the low $m$-modes ($m=2,\ 3$ or 4) dominate the power spectrum, and (2) the Fourier Amplitude in these modes (i.e., surface density contrast in structures) is $\sim$0.1 or larger.

{In total, GI-induced structures remain readily detectable ($\sim$ hours with ALMA) for the first few $10^4$ years after the system has emerged from its nascent envelope, when $q$ stays above $\sim$0.25. After this time ($\gtrsim 10^5$ years), detecting the substructure requires an integration time of $\sim$ days. Ultimately, the size scale of the substructure will drop below the resolution limit of ALMA, so even with increased integration time, resolving the substructure will not be possible.} \vspace{0.5cm}


\subsection{Temporal behaviour on dynamical timescales}
\label{subsec:dynamical}

We now turn to the behaviour of these systems on dynamical timescales. In section \ref{subsec:secular}, we assert that detection of spiral structure is easiest when the $m=2$ mode dominates the spectrum, so we examine if having a low amplitude in the $m=2$ mode will prevent detection, or alternatively, how else the power may be distributed in the mode spectrum so that detection is still possible. 


\begin{figure*}
\begin{tabular}{ccc}
$t=10^3$ years & $t=10^4$ years & $t=4\times 10^6$ years \\
$M_\mathrm{d}=0.5$ M$_\odot$ & $M_\mathrm{d}=0.25$ M$_\odot$ &  $M_\mathrm{d}=0.075$ M$_\odot$ \\
\includegraphics[width=0.33\linewidth]{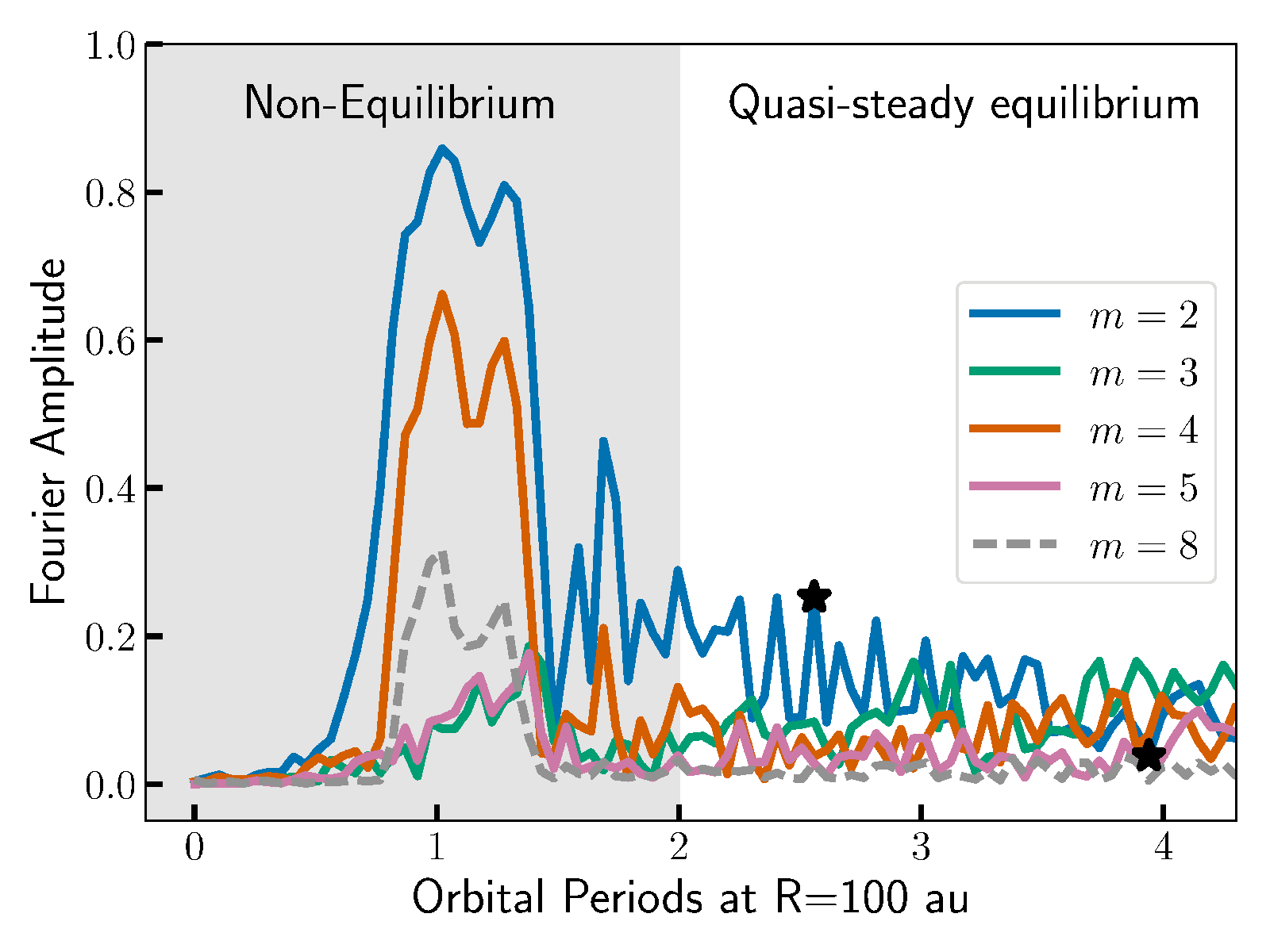} & \includegraphics[width=0.33\linewidth]{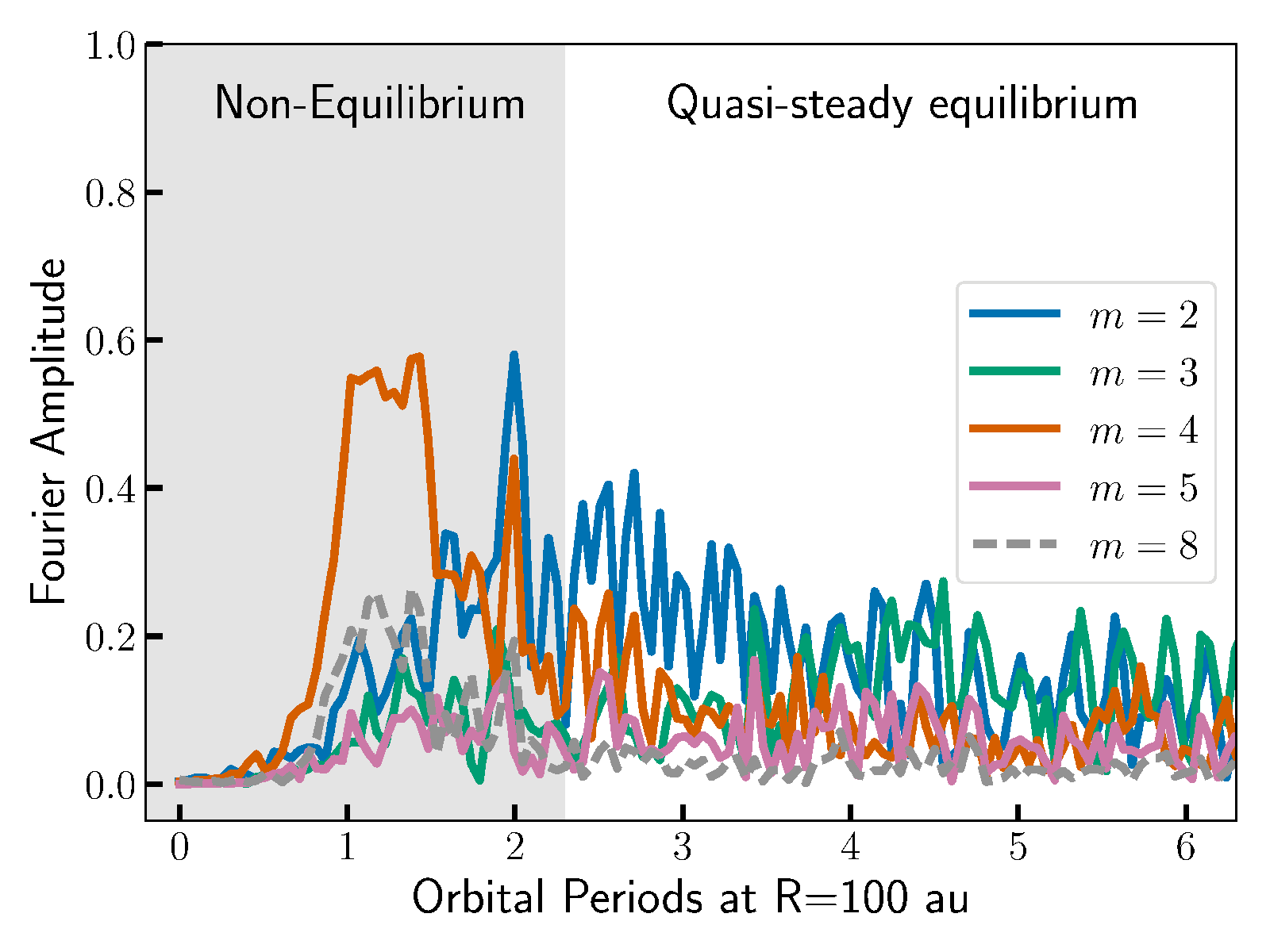} & \includegraphics[width=0.33\linewidth]{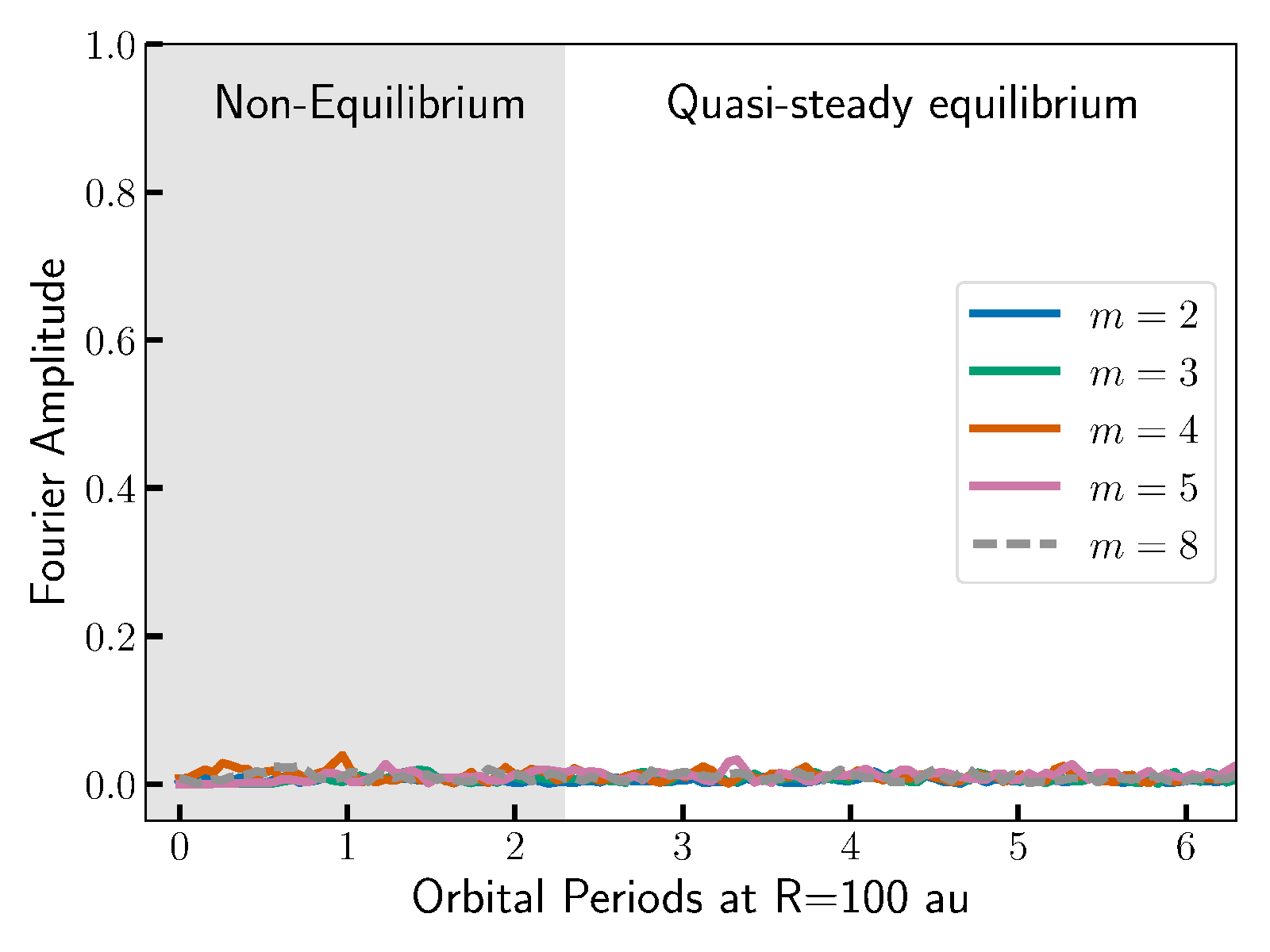}
\end{tabular}
\caption{
Fourier amplitudes of $m=2$, $m=4$ and $m=8$ modes for discs at $10^3$ years, $10^4$ years and $10^6$ years. Each plot should be considered separately.  For example, at $t=0$ the disc has undergone some evolution and infall from its nascent cloud, which increases the surface density such that large, transient spiral waves are produced until the system settles into a state of quasi-steady equilibrium. At later stages of the disc lifetime, there is very little power in the lower $m$-modes, which are more readily detectable by ALMA due to density contrast and larger spatial separation.\label{fig:fourierVtimedynamical}}
\end{figure*}

Figure \ref{fig:fourierVtimedynamical} shows the amplitude of the $m=2,3,4,5$ and 8 Fourier modes as a function of time for the $q=0.5, 0.25$ and 0.075 discs. At $t=0$, all particles in the disc are in exact Keplerian rotation. The more massive discs then undergo a period of violent relaxation where large, global $m=2$ modes rapidly redistribute angular momentum until the disc settles into a quasi-steady state. We select two points in time for the $q=0.5$ disc, shown in the leftmost panel of Figure \ref{fig:fourierVtimedynamical}, marked with black stars. The first point, at $\sim 2.5$ outer orbital periods, is when the $m=2$ Fourier amplitude is highest. The second point, at $\sim 4$ outer orbital periods, is when the $m=2$ Fourier amplitude is lowest during the quasi-steady state. We note, however, that in the $q=0.5$ case, even as power decreases in the $m=2$ mode, power can still be found in the $m=3-4$ modes. 

\begin{figure}
\vspace{0.5cm}
\hspace{1.7cm}A$_{(m=2)}$=0.25 \hspace{1.2cm} A$_{(m=2)}$=0.03 \\
\includegraphics[width=1.2\linewidth,trim={0 70 0 0}]{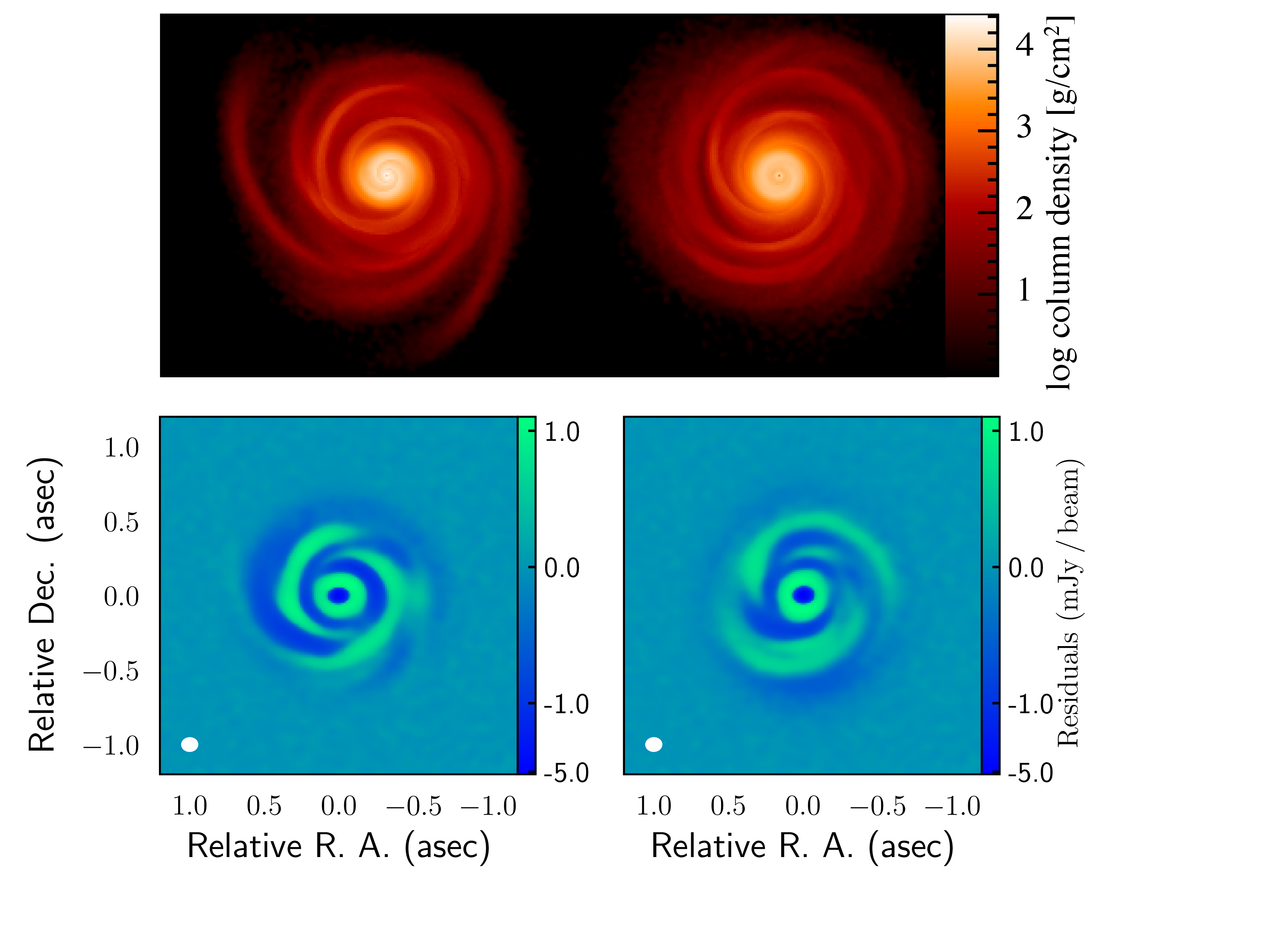} 
\caption{This figure shows two snapshots for the $q=0.5$ disc, when the Fourier amplitude of the $m=2$ mode is at its highest (left) and lowest  (right). Top row is surface density of the SPH simulation, bottom row is the ALMA residuals. We can see that even when the $m=2$ amplitude is low, there is sufficient power in adjacent modes such that non-axisymmetric structure is still detectable. \label{fig:mhighlow}}
\end{figure}

We generate synthetic ``residual'' images at these two epochs in Figure \ref{fig:mhighlow}. The top row shows the surface density structure of the discs, and the bottom row shows the ALMA residuals. In both cases, spiral structure is clearly detectable, despite the order of magnitude difference in the amplitude of the $m=2$ mode. \textit{We therefore conclude that the detectability of GI spirals has little dependence upon behaviour exhibited on the dynamical timescale of the system}. As long as the disc mass is high enough, there will be sufficient power in the low $m$-modes and the spirals will remain detectable, even if the amplitude oscillates between modes.



\subsection{Caveats}
\label{sec:caveats}
Interpreting disc observations is impossible without the use of numerical models, and generating these models are challenging. To properly model a system, we require full polychromatic radiation hydrodynamics, chemistry as well as dust dynamics and back reaction. To do all of these is not, at the time of writing, computationally possible \citep{haworthetal2016}. Although we use a state of the art radiative transfer approximation, we do not consider chemistry, and we assume that the dust and gas are well mixed. 


In a self-gravitating disc, spiral arms are able to trap grains of certain sizes. This trapping is most effective for grain sizes of $\sim 10-100$ cm in a typical self-gravitating disc \citep{riceetal2004}, which contribute orders of magnitude less emission at $\sim$ mm wavelengths due to their low mm opacity \citep{draine2006} when compared to mm grains.
Therefore, since trapping in the regime of interest for our parameters is expected to be small, we assume that the dust and gas are well-mixed. Future simulations with gas plus mm-sized dust are needed to quantify the effect.

\section{Summary and discussion}
We have performed a series of 3D global SPH simulations of a self-gravitating disc at different stages, with the disc masses taken from the 1D models of \citet{ricearmitage2009}. Essentially, this is equivalent to taking snapshots in time of a disc undergoing secular evolution.
The 1D models captured the long-term evolution, while the 3D simulations captured the detailed disc structure and dynamical, transient effects. We have performed synthetic ALMA observations on these simulations, and used the unsharped image masking method to highlight asymmetric structures.

Our main conclusion is that in isolated systems, where evolution is driven primarily by GI, high amplitude, symmetric two-armed spirals should be rare. Such arms only last for a few $\times10^{4}$ years after envelope dispersal, due to the rapid evolution of such massive discs. 
In these same systems, lower amplitude, multi-arm spirals can persist for much longer ($\sim$ Myr timescales). However, long integrations (i.e, up to $72$ hours on source) and high angular resolution (i.e., $\lesssim$0.05$\arcsec$) is needed to detect them.

A fundamental limitation of our investigation is that these are gas-only simulations, and we assume gas and dust are perfectly coupled. In reality,as discussed in section \ref{sec:caveats} there would be at least some dust enhancement due to trapping at local pressure maxima. 

It is worth noting that as the disc mass decreases, the temperature at which $Q\sim 1$ will also decrease. Therefore, for lower disc masses (i.e., older discs), even a modest amount of external irradiation can wash out the spiral structure. In the environment where these objects are found, it is therefore increasingly difficult to maintain spiral structure with age.

Our further conclusions are as follows:

\begin{enumerate}
\item{GI can produce detectable structure in the residuals of ALMA images taken with $\sim$ hour long integrations when $q=M_{\mathrm{disc}}/M_*\gtrsim0.25$ and $R\lesssim 100$ au. Fourier analysis shows that this corresponds to a minimum Fourier amplitude (i.e., substructure surface density contrast) of $\sim 0.1$ in low $m$-modes with $m\lesssim 4$.}
\item For an isolated system (i.e., insubstantial infall from envelope), the phase with readily detectable GI-induced structure lasts for a few $\times10^4$ years after the infall has ceased (i.e., disc accretion rate surpasses infall rate).
\item{After this phase, dedicated $\sim$day-long integrations are able to detect substructure in discs with $q\sim 0.175$, corresponding to $10^5$ years after the cessation of infall. As disc mass continues to decrease, eventually substructure starts to exist on length scales below that of the highest ALMA resolution so cannot be detected.}
\item Temporal variation of Fourier mode amplitudes on dynamical timescales does not affect whether ALMA is able to detect the spirals. 
\end{enumerate}

Among protoplanetary discs imaged in scattered light, $\sim10-20$\% have been found to host two-armed spirals \citep[and more with multi-armed spirals]{dongetal2018}. In section 1, we posed the question ``how likely is GI the dominant spiral-driving mechanism in protoplanetary discs, given its observability and the observed occurrence rate of spirals in discs?''

Our models show that for two-armed spirals observed in dust continuum emission to be caused by GI, the system must either still be embedded and receiving mass via infall so as to maintain a high disc-to-star mass ratio, or have emerged from its natal envelope within the last a few $\times\sim 10^4$ years, i.e., approximately 1\% of the typical age of discs. 

If the system is embedded in an envelope, then it is possible that infall from this envelope can drive power into the lower $m$-modes of the disc \citep{harsonoetal2011}, which would probably increase the ease with which these spirals would be detected. 

Without diving into the specifics of each individual disc, we conclude that gravitational instability is unlikely to be the dominate mechanism driving observed spiral arms to date. 

This work has shown that it is very difficult to detect spirals in GI discs in the continuum with ALMA, when the disc-to-star mass ratio drops below $q\sim 0.125$, which would typically correspond to times $t \gtrsim 1$ Myr. 
Previous work has shown that some of the conditions that we demonstrate here would also be true for NIR scattered light. Specifically, that the disc be massive ($q\gtrsim 0.25$), compact, and have a relatively high accretion rate ($\gtrsim 10^{-6}$ M$_\odot$ yr$^{-1}$) \citep{donghallrice2015} in order to drive low-$m$ spirals.

However, even a very small amount of remaining nascent envelope can obscure the disc at NIR wavelengths. To clearly image the surface of the disc requires that there is virtually no envelope present, which probably requires $q\lesssim 0.1$. If such discs are imaged in NIR scattered light they should display a high number of spiral arms, since $m\sim 1/q$. Therefore, it is unlikely that GI can explain objects such as MWC 758 \citep{benistyetal2015} or SAO 206462 \citep{garufietal2013,stolkeretal2016}, since these objects show prominent $\sim 2$-armed spirals. 

GI may, however, offer an explanation for systems that display multiple, weaker spirals arms in scattered light, such as AB Aur \citep{hashimotoetal2011}. The measured disc mass of AB Aur from mm continuum observations is $\sim 20$ M$_J$ \citep{henningetal1998}, placing the disc-to-star mass ratio at $q\sim 0.01$. This is about 5-10 times fewer than the lower limit necessary for GI to be active. However, disc masses inferred from mm dust observations have many uncertainties. If grain growth has occurred, which may be reasonably likely in a system that is $\sim$ Myr old, then the disc is expected to be optically thick out to $\sim 3$ mm wavelengths. This can result in the underestimation of disc mass by an order of magnitude \citep{forganrice2013,dunhametal2014,evansetal2017,galvanmadridetal2018}. If this is the case for a system such as AB Aur, then the system may be self-gravitating.   

Recently, it has come to light that protoplanetary discs appear not to be massive enough to form the known exoplanet population \citep{manaraetal2018}. Either discs are being continually replenished from their environment (an unseen envelope), or cores of planets form very rapidly (between 0.1 and 1 Myr) and a large amount of gas is expelled shortly after their formation. If the ringed structure of systems such as HL Tau \citep{hltau} is due to planet formation \citep{dipierrohltau}, then the latter is certainly possible, although no mechanism is known that would remove so much mass from the system on such a short timescale, which would only dominate after the formation of planetary cores. However, this problem is solved, and it does not challenge the current planet formation paradigm, if it is simply the case that disc masses are being systematically underestimated, as we have discussed above.

Although the work presented in this paper does not focus on fragmentation due to GI, it is worth noting that the work of \citet{manaraetal2018} finds two distinct populations for single and multiple exoplanetary systems, which, they suggest, may point to a different formation mechanism for single exoplanets. Simply put, single exoplanets around low-mass stars can have masses that are comparable to their host star mass, which is never observed in multi-planet systems. This seems to be consistent with the current understanding of planet formation through GI, that these objects are, essentially, failed companion stars \citep{kratteretal2010b}.


Previous numerical investigations have found that fragmentation occurs in GI discs that extend beyond $R\gtrsim 50-100$ au \citep{rafikov2005,matznerlevin2005,whitworthstamatellos2006,clarke2009,clarkelodato2009,kratteretal2010b,forganrice2011,halletal2016,halletal2017,halletal2018}. 

Therefore, if a disc has spiral structure that extends beyond $\sim 100$ au, it is unlikely to be caused by GI. Although irradiation beyond these radii reduces the local effective gravitational stress, and, therefore, the amplitude of the spiral arms (see, e.g. \citealt{halletal2016}), it does not prevent fragmentation \citep{riceetal2011}. As irradiation increases, the disc behaves more like an isothermal system. As such, even large amplitude spirals do little to dissipate thermal energy and redress the thermal balance\citep{krattermurrayclay2011}, resulting in non-linear growth of the spirals and ultimately fragmentation.

The results presented here are consistent with the results of \citet{halletal2016}, who found that detecting signatures of disc self-gravity with ALMA required the disc to exist in a very narrow region of parameter space, where the spiral wave amplitudes are large enough to produce detectable features, but not so large as to cause the disc to fragment. In this work, we have examined the region of parameter space where the semi-analytical approach of \citet{halletal2016} would not have been valid, i.e., we have simulated global angular momentum transport by global (loosely wound, low $m$) spiral arms, rather than only considering the local regime (tightly wound spiral arms, high $m$) that can be described by a semi-analytical model. Essentially, in \citet{halletal2016}, it was found that it is difficult to detect spiral arms caused by GI in the local regime ($q\lesssim 0.25$), and we again find this result in this work.

Until recently, the discs around Elias 2-27 \citep{perezetal2016} and MWC 758 \citep{dongliuetal2018} were the only confirmed cases of $m=2$ spiral arms in protoplanetary discs imaged in mm continuum emission.  However, the DSHARP (Disk Substructures at High Angular Resolution Project) ALMA survey  \citep{dsharp1} has revealed four new instances of spiral arm structure in 1.25 mm emission in protoplanetary discs, around the systems IM Lup, WaOph 6, HT Lup A and AS 205 N, as well as a more high-resolution observation of Elias 2-27. HT Lup A and AS 205 N are multi-disc systems, so it is likely that interactions between these components has given rise to the spiral structures present \citep{kurtovicdsharp}.

The Elias 2-27, IM Lup and WaOph 6 systems, however, have no known companions. Although it may be possible that the $m=2$ spiral structure present in these systems is due to GI, it has traditionally been thought that such Class II systems would be too low-mass to be susceptible to gravitational instability. This is compounded by the fact that measurements of the Toomre parameter for Elias 2-27 and IM Lup indicate that these discs should be stable to GI \citep{perezetal2016,cleevesetal2016}. Furthermore, the recent high-resolution observations of these systems have revealed annular substructure in all three discs \citep{dsharpspirals}, in addition to the $m=2$ spiral arm pattern. It is difficult to explain the coexistence of spirals and annuli together with GI alone, suggesting that either GI is not acting, or it is present in conjunction with another mechanism responsible for the annular substructure.

Finally, the age estimates for Elias 2-27, Im Lup and WaOph 6 are 0.8 Myr, 0.5 Myr and 0.3 Myr respectively  \citep{luhmanrieke1999,alcalaetal2017,eisneretal2005}. The work we have presented here has shown that it is unlikley that $m=2$ spiral arms caused by GI persist at these ages.  Systems with a higher number of weak GI-induced spiral arms can persist for far longer, $\sim 10^6$ years. However, these systems are far more difficult to detect with ALMA because the low contrast requires high sensitivity, and the high $m$-modes demand high angular resolution. It is yet to be seen whether the required observing conditions and integration times are realistic or not.






\acknowledgments
CH is a Winton Fellow and this research has been supported by Winton Philanthropies. This research used the ALICE2 High Performance Computing Facility at the University of Leicester. This research also used the DiRAC DIaL (Data Intensive at Leicester) facility.
We would like to thank Daniel Price for his publicly available SPH plotting code \texttt{SPLASH} \citep{splash}, which we have made use of in this paper. This project has received funding from the European Research Council (ERC) under the European Union's Horizon 2020 research and innovation programme (grant agreement No 681601). TJH acknowledges funding from Exeter's STFC Consolidated Grant (ST/M00127X/1).

\bibliographystyle{aasjournal}
\bibliography{bib}{}

\end{document}